\documentclass[12pt]{iopart}
\usepackage{graphics}

\usepackage{iopams}  
\begin{document}

\title[Controllable diffusion of cold atoms]{Controllable diffusion of cold atoms in a harmonically driven and tilted optical lattice: Decoherence by spontaneous emission}

\author{Navinder Singh}

\address{Institute of Physics, Bhubaneswar-751005, India}
\ead{navinder@iopb.res.in}
\begin{abstract}
We have studied some transport properties of cold atoms in an accelerated optical lattice in the presence of decohering effects due to spontaneous emission. One new feature added is the effect of an external AC drive. As a result we obtain a tunable diffusion coefficient and it's nonlinear enhancement with increasing drive amplitude. We report an interesting maximum diffusion condition.
\end{abstract}

\vspace{2pc}
\maketitle

\section{Introduction}
\label{intro}
The seminal review paper\cite{stig} of Stig Stenholm on the theory of laser cooling gave a boost to cold atom physics and optical laser technology\cite{ad94,jess}. Cold atoms on optical lattices constitute {\em clean} quantum systems as compared to solid state systems due to very less scattering and decohering effects, and provide a paradigmatic model to test some novel quantum phenomena predicted half century back for metallic systems\cite{niu97,bloch28}. The experimental advancements has further provided  much more deeper understanding in fundamental quantum statistical laws, new insights in quantum computing technologies, quantum phase transitions (superfluid to Mott-insulating phase) in the regime of strong correlations\cite{greiner}, and Bose-Einstein condensation on optical lattices\cite{zoller,mac,meacher,bloch08,oli06}.

In the present contribution, we consider single atom transport (in one dimension) on a tilted optical lattice in 
the presence of an external AC field. To fix ideas, consider the physical situation of an atom interacting with laser field. We have a single atom A, with an excited state $|e\rangle$ and a ground state $|g\rangle$ separated by an energy interval $E_e - E_g = \hbar\omega_A$. The atom Hamiltonian is $H_A = p^2/2m + \hbar\omega_A|e\rangle\langle e|$ (with $E_g = 0$). The atom is subjected to a classical laser field with electric field ${\bf E(x,t)} = \epsilon E(x)e^{-i\omega_L t}$, where $\omega_L$ is the laser frequency and ${\bf \epsilon}$ is the polarization vector of the laser. If the amplitude of the electric field E(x) is varying slowly in space $x$ compared to the size of the atom, the   atom-field interaction can be described in dipole approximation, i.e., by the coupling $= -\mu {\bf E(x,t)}$, where $\mu$ is the atomic dipole moment. We assume that laser is far detuned from any optical transition in the atom. In this simplified picture one uses perturbation theory and eliminate internal atomic states from the dynamics to obtain an effective potential $V(x)$. The atom Hamiltonian thus becomes  $H_A = P^2/2m + V(x)$, with $V(x) = |\Omega(x)|^2/4(\omega_L-\omega_A)$. The term $\Omega(x) = -2E(x)\langle e|\mu \epsilon|g\rangle$ is called the Rabi frequency which drive the transitions between the two atomic levels. Now consider that the atom is in a non-resonant standing light $( E(x,t)=2\epsilon E_0 \cos k_Lx\cos \omega_L t=\epsilon E_0[\cos(\omega_L t-k_Lx)+\cos(\omega_L t+k_Lx)],~two~~counter~propagating~waves)$. This constitutes 'The Basic' optical periodic potential $U(x) = U_0\cos^2(k_L x)$, which can be accelerated by frequency chirping. Lattice Potential is tilted in the reference frame of the atom (co-moving frame), and the atomic motion is governed by quantum mechanics of a particle on a periodic lattice (Figure 1). When there is no acceleration (no tilt), an initially localized wavepacket will spread through resonant Bloch tunneling and become delocalized. But in a  tilted lattice (optical lattice being accelerated analogously electron and crystalline lattice is in an external constant electric field) atoms can remain localized due to suppression of Bloch tunneling. They exhibit novel quantum phenomenon of Bloch oscillations due to the repeated Bragg scattering\cite{ben96}. In the case of electrons in metals  this corresponds to an induced AC current with applied DC voltage across the sample. But in  usual practice this purely quantum effect is overshadowed by scattering processes and we obtain ohomic DC current. The second interesting effect of potential tilt is that the Bloch bands are broken up into Wannier-Stark (WS) Ladders of states\cite{wan60,fuku}. The level spacing between two nearby levels in the ladder is given by  $eEd$ ($e$ is the electron charge, $E$ is the applied constant electric field and $d$ is the lattice spacing, for the case of optical lattice, level spacing $= Fd = mad $, $a$ is the imparted acceleration, and $d=\pi/k_L$ is the period of optical potential). As the tilt of potential per lattice spacing becomes comparable to well depth, a new interband tunneling process called Landau-Zener tunneling becomes important which is a natural extension of the stark effect for a single atom.

Now the above stated phenomena are purely quantum in nature. But the relaxation processes are natural. No system is an ideally decoupled system for all space and time scales. In case of cold atoms on optical lattices the main relaxation process is the spontaneous emission of photons by excited atoms. Since photons has finite momentum, it's generation gives a recoil kick to the atom(mechanical effects of light). These relaxation processes decoher pure quantum effects. The decay of Bloch oscillations and the diffusive spreading of the atoms is shown to cause by spontaneous emission\cite{kol02}. Interestingly, the presence of relaxation processes are actually important for practical purposes, like ohomic current across metals(Joule heating effects) and atomic current across an optical lattice etc.\cite{pono05}.

In the present contribution we consider the effect of external AC drive on the quantum transport of cold atoms on tilted optical lattice, and the effect of spontaneous emission that cause decoherence, which is essential for diffusive motion. We have obtained a tunable diffusion co-efficient and it's nonlinear enhancement with increasing drive amplitude and we also report a novel maximum diffusion condition. The analytical results obtained by us correctly specialize to the exact results known in the limit of zero drive and zero-bias. Also, in addition to cold atoms, the results obtained are applicable to experimentally realizable super lattice hetero-structures that support the Stark-Wannier (SW) ladder states in the presence of a strong longitudinal electric-field bias\cite{niu97,wan60,mg02,nav05}. As is well known, a strong field normal to the superlattice planes can break up the extended Bloch-like band continuum into energetically well resolved states localized in the potential wells. The stronger the biasing field the more localized are the SW state\cite{wan60,fuku}. 

This paper is organized as follows. In section 2 we introduce the model Hamiltonian and the simplified master equation for the present system. In section 2.1 we consider  the simple case of no acceleration and no external AC drive, and in section 2.2 tilted lattice in the presence of an external AC drive. We end with brief discussion of results.
\begin{figure}
\begin{center}
\scalebox{0.6}[0.6]{\includegraphics{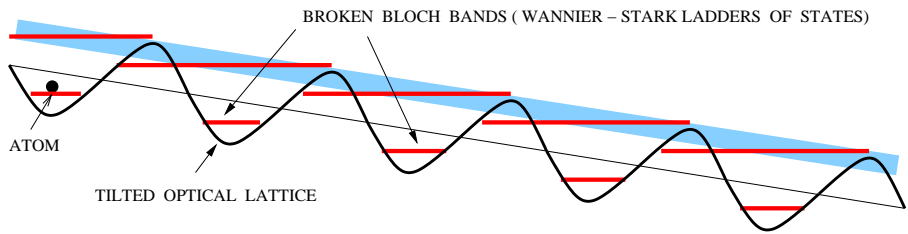}}
\end{center}
\caption{An atom in an accelerated optical periodic potential (in co-moving frame of reference it is a tilted or washboard potential)}
\label{fig:1}
\end{figure}
\section{Model Hamiltonian and Simplified Master equation}
\label{sec:1}
We have three interacting physical systems (1) the atom (2) the laser field, and (3) the vacuum. Up to this point we have not considered atom-vacuum field coupling, which is responsible for spontaneous emission of photons by the excited atom. The random recoil kicks (mechanical effects of spontaneously emitted photon momentum) cause decoherence of atomic motion, which is the main object under study. Spontaneous emission is characterized by the natural width of the excited state or the radiative life-time of the state ($=1/\gamma$). If we are interested in very short interaction times ($t\ll 1/\gamma$), we can neglect spontaneous emission, and the evolution of atom-laser system is described by Schrodinger equation. But for long interaction times $(t\gg 1/\gamma)$, due to the presence of several spontaneous emissions, one cannot use Schrodinger equation. In this situation one uses system-environment approach i.e., the reduced atomic evolution (traced over infinitely many vacuum field (environment) degrees of freedom) is then given by a master equation. In the present situation of far detuning $(\delta_0 = \omega_L - \omega_0 \gg \Omega(atomic~Rabi~frequency)$ ,the time evolution of the  reduced density matrix $\rho$ of the atomic motion along the $x$ direction is given by a simplified master equation\cite{kol02} (detailed theory is given in\cite{gra96,goet96,mg02,ad94,let81}): 
\begin{equation}
\frac{\partial\rho}{\partial t} =-\frac{i}{\hbar}[H,\rho] -\frac{\gamma}{2}\int du p(u)[L_u^{\dag}L_u\rho +\rho L_u^{\dag}L_u - 2L_u\rho L_u^{\dag}]\,\,,
\end{equation}
Where $\gamma = \gamma_0\frac{\Omega^2}{\delta_0^2}$($\gamma_0$ is the inverse radiative life time of the excited state), $\Omega$ is the atomic Rabi frequency, and $\delta_0$ is static detuning. The first term on the RHS gives the unitary evolution, while the second term gives the non-unitary (incoherent) evolution causing the initially pure density matrix  ($\rho = \rho^2$ at $t=0$) to become mixed ($\rho \neq \rho^2$ for $t > 0$). Here $p(u)$ is the angle distribution of the spontaneously  emitted photons, and for linearly polarized light, $p(u) \simeq 1/2$. The operator $L_u$ is the projection of photon recoil operator along the atomic direction of motion. The photon recoil operator represents the coupling of the internal atomic dynamics (decay processes) to the external atomic motion(center of mass motion). In dipole and rotating-wave approximation, 
\begin{equation}
L_u = cos(k_L x) e^{i u k_L x}\,\,\, |u| \leq 1
\end{equation}
The physical effect of the photon recoil operator is to randomly change the atomic quasimomentum. Initial delta function distribution of atomic quasimomentum will be smeared over the entire Brillouin zone and cause the decay of Bloch oscillations as observed in\cite{kol02}.
Our approach is based on one band tight binding hamiltonians similar to\cite{kol02}(new feature added is the AC drive, section 2.2). In matrix element notation, our system Hamiltonian and Recoil operator is
\begin{equation}
H_{mn} = -\frac{V}{2}[\delta_{m,n+1} +\delta_{m,n-1}] + F d m\delta_{mn}
\end{equation}
\begin{equation}
L_{mn} = (-1)^m e^{i\pi u m}\delta_{mn}
\end{equation}

Here, $V(>0)$ is the transfer matrix element between nearby states, $F$ is the inertial force acting on the atom, $d = \pi/k_L$ is the period of optical potential, and $m$ is the localized Wannier function $|m\rangle$ associated with the $l$th well of the optical periodic potential. In the following sections we will proceed step by step, starting with un-accelerated potential.
\subsection{No acceleration and no AC Drive}
\label{sec:2}
\begin{figure}
\scalebox{.8}{\includegraphics{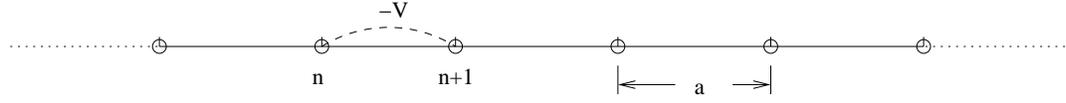}}
\caption{No acceleration case, all states having same energy.}
\label{fig:2}
\end{figure}
We begin by considering first the simplest case of quantum motion of an atom moving on an optical lattice (Figure 2) under a tight-binding one-band Hamiltonian
\begin{equation}
H^0 = - (V/2)\sum_{l} \left(\mid l\rangle \langle l+ 1\mid + \mid l+1\rangle\langle l \mid 
\right) 
\label{1}
\end{equation}
where (-$V$) is the nearest-neighbor transfer matrix element, and the sum is over the $N$ sites with $N$ taken to be infinite. The effect of spontaneous emission, namely the incoherence, will be introduced through recoil operator defined in equation 4. The reduced density matrix $\rho$ for the particle then obeys the evolution master equation (equation 1). In terms of  matrix elements, we have
\begin{equation}
\frac{\partial \rho_{mn}}{\partial t}=-\frac{i V}{2\hbar}[\rho_{m,n+1} + \rho_{m,n-1} - \rho_{m+1,n} - \rho_{m-1,n}] - \gamma \rho_{mn}[1-\delta_{mn}].
\label{3}
\end{equation}
With the initial condition,
\begin{equation}
\rho_{mn}(t = 0) = \delta_{m0}\delta_{n0}.
\label{4}
\end{equation}
In Fourier space, with $\beta = -\frac{V}{\hbar}$. 
\begin{eqnarray}
&&\frac{\partial}{\partial t} \sum_{m,n}\rho_{m,n} e^{-i m k_1} e^{i n k_2}=
\frac{i\beta}{2}\left[\sum_{m,n}\rho_{m,n+1} e^{-i m k_1} e^{i (n+1) k_2}e^{-i k_2}+ . . .\right]\nonumber\\
&&-\gamma\sum_{m,n}\rho_{m,n} e^{-i m k_1}e^{i n k_2} + \gamma\sum_{m,n}\delta_{m,n} e^{-i m k_1}e^{i n k_2},
\label{5}
\end{eqnarray}
with
\begin{equation}
\tilde{\rho}(k_1,k_2,t) = \sum_{m,n}\rho_{m,n} e^{-i m k_1}e^{i n k_2},
\delta_{m,n}=\frac{1}{2\pi}\int_{-\pi}^{\pi}e^{i(m-n)q} dq.
\label{6}
\end{equation}
We get
\begin{eqnarray}
&&\frac{\partial}{\partial t}\tilde{\rho}(k_1,k_2,t)=[i\beta(\cos k_2-\cos k_1)-\gamma]\tilde{\rho}(k_1,k_2,t)\nonumber\\ 
&&+\frac{\gamma}{2\pi}\int_{-\pi}^{\pi}\tilde{\rho}(k_1-q,k_2-q,t)dq.
\label{7}
\end{eqnarray}
Defining center of mass and relative wave-vectors as $p=(k_1+k_2)/2,\,\,u=k_1-k_2$ and writing $\tilde{\rho}(k_1,k_2,t)\equiv \rho(p,u,t)$ we have
\begin{equation}
\frac{\partial}{\partial t}\rho(p,u,t)= [2i\beta\sin p \sin(u/2)-\gamma]\rho(p,u,t)
+\frac{\gamma}{2\pi}\int_{-\pi}^{\pi}\rho(p-q,u,t)dq.
\label{8}
\end{equation}
We define further reduced density matrix by
\begin{equation}
\chi(u,t)=\frac{1}{2\pi}\int_{-\pi}^{\pi}\rho(p-q,u,t)dq =\frac{1}{2\pi}\int_{-\pi}^{\pi}
\rho(q,u,t)dq.
\label{9}
\end{equation}
As the dimensions of $\beta$ and $\gamma$ are $(time)^{-1}$, we define the dimensionless
parameters as $\tau = t\beta$ and $\Gamma = \frac{\gamma}{\beta}$,
and defining
\begin{equation}
\phi(p,u) = 2i\sin p \sin(u/2)-\Gamma.
\label{10}
\end{equation}
So with this, the evolution equation becomes
\begin{equation}
\frac{\partial}{\partial \tau}\rho(p,u,\tau)= \phi(p,u)\rho(p,u,\tau)+\Gamma\chi(u,\tau).
\label{11}
\end{equation}
The detailed calculation of mean-squared displacement from the above equation is given in Appendix A. We finally obtain
\begin{equation}
\langle{x^2}(\tau)\rangle = -\frac{1}{\Gamma^2}+\frac{1}{\Gamma}\tau+\frac{1}{\Gamma^2}e^{-\Gamma\tau}.
\label{21}
\end{equation}
Recalling, $\tau = t\beta$ and $\Gamma = \gamma/\beta$, where $\beta=V/\hbar$, 
$\gamma = \gamma_0\frac{\Omega^2}{\delta_0^2} $, with this
\begin{equation}
\langle{x^2}(t)\rangle = -\frac{\beta^2}{\gamma^2}+\frac{\beta^2}{\gamma}t + 
\frac{\beta^2}{\gamma^2}e^{-\gamma t}\;\;,
\label{22}
\end{equation}
which reduces to the classical case in the large time ($t \gg 1/\gamma$) limit as 
$\left<x^2(t)\right>\,\,\sim\,\, ({\beta^2}/{\gamma}) t $ giving the diffusion coefficient 
\begin{equation}
D=\frac{\beta^2}{2\gamma} = \frac{V^2}{2\gamma_0\hbar^2}\left[\frac{\omega_L-\omega_0}{\Omega}\right]^2
\end{equation}
In the small time limit it goes ballistically as $t^2$ as expected, while the 
mean displacement $\left<x(t)\right>$ is zero ({\em no atomic current on an un-accelerated lattice}).
\subsection{Accelerated lattice in an external AC drive}
When an acceleration is imparted to the lattice, the Block bands are broken up into Wannier-Stark Ladder (WSL) of states with level spacing equal to $Fd$, in other words lattice Hamiltonian 
has a systematic bias. There is a constant energy mismatch  between the successive site energies (Figure 3). Now consider that this system of discrete energy states is present in an external AC laser field. We describe this physical picture by a tight-binding one-band Hamiltonian
\begin{figure}
\begin{center}
\scalebox{1.2}[1]{\includegraphics{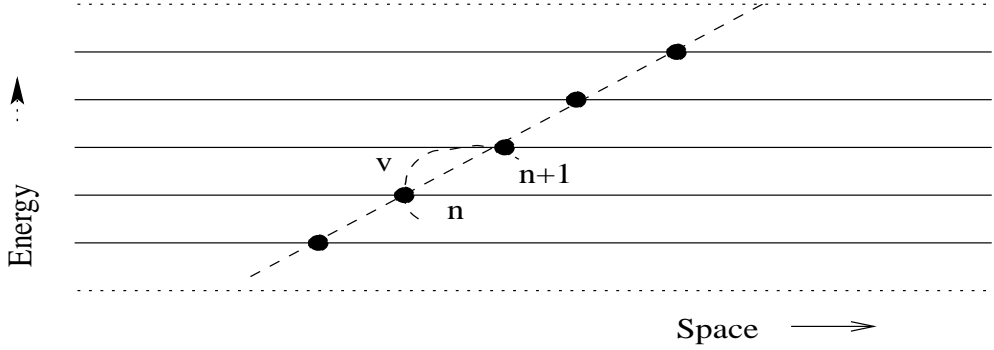}}
\end{center}
\caption{System with discrete energy levels (WS Ladder) with level spacing $F d$}
\label{fig:3}
\end{figure}
\begin{equation}
H^{\omega} = -E_0 \cos \omega t \sum_l[|l\rangle\langle l+1| + |l+1\rangle\langle l |]+
\sum_l \alpha l |l\rangle\langle l|,
\label{53}
\end{equation}
The wavelength of external standing wave AC field is assumed to be much longer than lattice period, so as to subtend a relatively strong dipolar matrix-element between the neighboring WS states of the ladder. So, in the Hamiltonian, the term $E_0 \cos \omega t$ (time dependent drive of amplitude $E_0$ and circular 
frequency $\omega$) acts as a nearest-neighbour transfer matrix element. It may be noted, that in the limit $\omega = 0$, this simulates the usual transfer matrix element $-V = -E_0$. As before, the time evolution of the reduced density matrix (in matrix elements) is given by,
\begin{eqnarray}
&&\frac{\partial\rho_{mn}}{\partial t}=\frac{iE_0}{\hbar}[\rho_{m+1,n}+\rho_{m-1,n}-\rho_{m,n-1}-\rho_{m,n+1}]
-i\frac{\alpha}{\hbar}(m\rho_{mn}-n\rho_{mn})\nonumber\\
&&-\gamma[1-\delta_{mn}]\rho_{mn}
\label{55}
\end{eqnarray}
The quantities $\frac{E_0}{\hbar},\;\;\frac{\alpha}{\hbar}$, and $\gamma$ have a dimension of $time^{-1}$. We define $t_0=\frac{\hbar}{E_0}$, $\delta$(dimensionless acceleration) $=\frac{\alpha}{E_0}$, $\Gamma$(dimensionless decohering factor)$ = t_0\gamma$, and $\tau$ (dimensionless time)$ = \frac{t}{t_0}$. With this we have
\begin{eqnarray}
&&\frac{\partial\rho_{mn}}{\partial \tau} = i\cos\omega\tau[\rho_{m+1,n}+\rho_{m-1,n}-\rho_{m,n-1}-\rho_{m,n+1}]\nonumber\\
&&-i\delta(m\rho_{mn}-n\rho_{mn})-\Gamma[1-\delta_{mn}]\rho_{mn}
\label{56}
\end{eqnarray}
After applying the rotating wave approximation with $\rho_{mn} = \bar{\rho}_{mn}$
$e^{- i\delta (m-n)\tau}$, the evolution of reduced density matrix is,
\begin{eqnarray}
&&\frac{\partial \bar{\rho}_{mn}}{\partial \tau}=\frac{i}{2} [e^{i(\theta-\delta)\tau}
[\bar{\rho}_{m+1,n} - \bar{\rho}_{m,n-1}]\nonumber\\
&& + e^{-i(\theta-\delta)\tau}[\bar{\rho}_{m-1,n}-\bar{\rho}_{m,n+1}]]- \Gamma \bar{\rho}_{mn}[1-\delta_{mn}].
\label{57}
\end{eqnarray}
Here, we have defined $\theta = \omega t_0$,  $\tau = t/t_0$, ( $t_0 = \hbar/E_0$), and $\Delta = \theta - \delta$ as the detuning between drive frequency $\omega$ and scaled energy level spacing $\delta$. Note that
\begin{eqnarray}
&&\bar{\tilde{\rho}}(k_1,k_2,\tau) = \sum_{m,n}\bar{\rho}_{mn}(\tau)e^{- i m k_1}e^{i n k_2}
\nonumber\\
&&\bar{\rho}_{m,n+1} \rightarrow e^{-i k_2}\bar{\tilde{\rho}}(k_1,k_2,\tau)\;,\;\bar{\rho}_{m+1,n} \rightarrow e^{i k_1}\bar{\tilde{\rho}}(k_1,k_2,\tau)\nonumber\\
&&\bar{\rho}_{m,n-1} \rightarrow e^{i k_2}\bar{\tilde{\rho}}(k_1,k_2,\tau)\;,\;\bar{\rho}_{m-1,n} \rightarrow e^{-i k_1}\bar{\tilde{\rho}}(k_1,k_2,\tau).\nonumber\\
\label{58}
\end{eqnarray}
\begin{eqnarray}
&&\frac{\partial \bar{\tilde{\rho}}(k_1,k_2,\tau)}{\partial \tau} = (i[\cos (k_1+\Delta \tau) 
-\cos (k_2+\Delta\tau)]- \Gamma )\nonumber\\
&&\bar{\tilde{\rho}}(k_1,k_2,\tau)+
\frac{\Gamma}{2\pi}\int_{-\pi}^{\pi}\bar{\tilde{\rho}}(k_1-q,k_2-q,\tau)dq.
\label{59}
\end{eqnarray}
Performing similar co-ordinate transformations $p=(k_1+k_2)/2\;,\;u = k_2-k_1$ and defining $\bar{\tilde{\rho}}(k_1,k_2,\tau) \equiv \varrho(p,u,\tau)$, we have
\begin{eqnarray}
&&\frac{\partial \varrho(p,u,\tau)}{\partial \tau} = [2i \sin(p+\Delta\tau)\sin(u/2)-\Gamma]
\times\varrho(p,u,\tau)\nonumber\\
&&+\frac{\Gamma}{2\pi}\int_{-\pi}^{\pi}\varrho(p-q,u,\tau)dq.
\label{60}
\end{eqnarray}
The solution of first order P.D.E. (equation 24) is given in appendix B, we finally obtain
\begin{eqnarray}
&&\left<x^2(\tau)\right> =\frac{\Gamma}{\Gamma ^2 + \Delta ^2}\tau+\left[\frac{\Delta ^2-
\Gamma^2}{(\Delta ^2+\Gamma^2)^2}\right]\left \{1-e^{-\Gamma \tau}\cos\Delta \tau\right\}\nonumber\\
&&-\frac{2\Gamma\Delta}{(\Delta ^2+\Gamma^2)^2}e^{-\Gamma \tau}\sin\Delta\tau.
\label{73}
\end{eqnarray}
Noting that, $\Gamma$(dimensionless decohering factor) $ = \gamma\frac{\hbar}{E_0}$, $\tau$(scaled time)$ = t\frac{E_0}{\hbar}$, 
and $\Delta$(dimensionless detuning)$ = \frac{\hbar\omega-\alpha}{E_0}$.
The above equation (equation (25))is an important result of the present work.
The mean squared displacement from above equation is plotted in Figures 4 to 7.
\begin{figure}
\begin{center}
\scalebox{1.3}[1.3]{\includegraphics{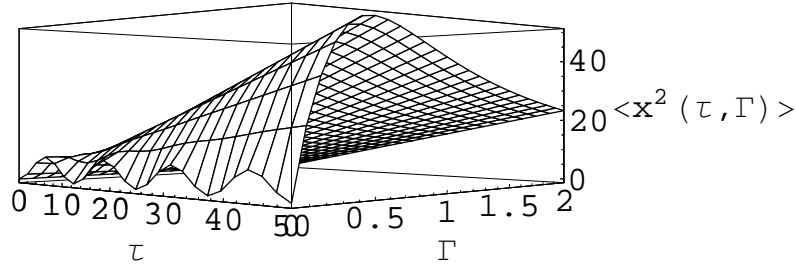}}
\end{center}
\caption{Plot of mean-squared displacement $\langle x^2(\tau,\Gamma)\rangle$ vs scaled time $\tau$ and dimensionless decohering factor $\Gamma = \gamma_0\frac{\hbar\Omega^2}{E_0\delta_0^2}$ (proportional to inverse  of radiative lifetime of excited state). One can clearly identify 'a peculiar transition point' near $\Gamma = 0.5$, below the transition point $\langle x^2(\tau,\Gamma)\rangle$  increases with increasing $\Gamma$ and above the transition point mean-squared  displacement decreases with increasing $\Gamma$. Oscillations smooth away as $\Gamma$ increases. Here $\Delta = 0.5$}
\label{fig:9}
\end{figure}
\begin{figure}
\begin{center}
\scalebox{1}[1]{\includegraphics{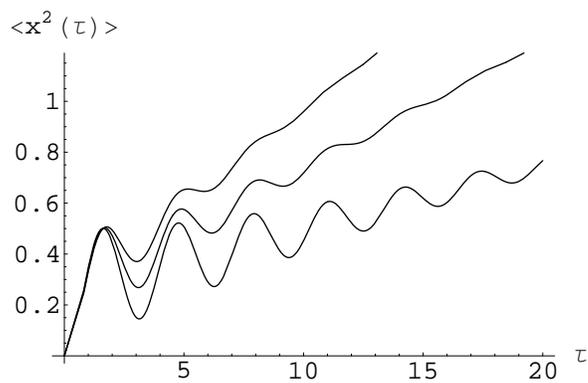}}
\end{center}
\caption{Enhancement of diffusion with increase in $\Gamma$. The top most curve is for $\Gamma=0.3$, central for $\Gamma = 0.2$, 
and lowest for $\Gamma=0.1$ (for $\Gamma < 0.5$). With constant detuning parameter $\Delta=2$. As $\Gamma$ decreases, the oscillations in the mean-squared displacement increases, but after a long time oscillations vanish and the mean-squared displacement goes linearly with time as it should. Oscillations are a signature of WSL due repeated reflections from nearby states.}
\label{fig:7}
\end{figure}
\begin{figure}
\begin{center}
\scalebox{1}[1]{\includegraphics{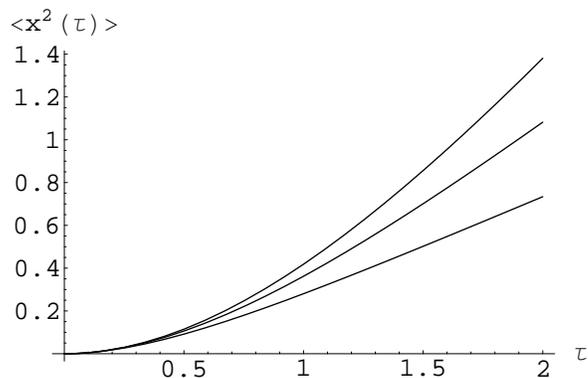}}
\end{center}
\caption{Suppression of diffusion due to increase in $\Gamma$ (for $\Gamma>0.5$). The top most curve is for lowest damping constant $\Gamma=0.5$, central for 
$\Gamma = 1$, and lowest for $\Gamma=2$. With constant detuning parameter $\Delta=0.5$.}
\label{fig:8}
\end{figure}
\begin{figure}
\begin{center}
\scalebox{1}[1]{\includegraphics{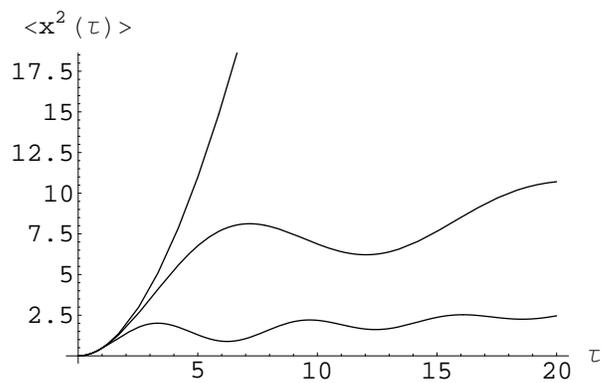}}
\end{center}
\caption{The effect of detuning $\Delta\sim \frac{\hbar\omega-\alpha}{E_0}$ on mean-squared displacement. The top 
most curve is for resonance case, no detuning $\Delta = 0$, central for $\Delta = 0.5$, and
lowest for $\Delta=1$. With constant $\Gamma=0.08$. As the detuning
goes up, the oscillations in the mean-squared displacement increases, but we have the same expected evolution, short time, $\tau^2$ rise, then oscillations, and after a long time oscillations vanish and the mean-squared displacement goes linearly with time.}
\label{fig:10}
\end{figure}
For two special cases of interest in long time limit, equation (25) gives:

(A) On-resonance, i.e., (detuning) $\Delta = \frac{\hbar\omega-\alpha}{E_0} = 0$,
\begin{equation}
\left<x^2(t)\right> = \frac{E_0^2}{\hbar^2\gamma}t\;,\;\; (diffusive).
\label{74} 
\end{equation}
(B) Off-resonance (finite detuning) $\Delta \neq 0$,
\begin{equation}
\left<x^2(t)\right> =\frac{E_0^2\gamma}{\hbar^2\gamma ^2 +
(\hbar\omega-\alpha)^2}t\;,\;(diffusive-controllable)
\label{75}
\end{equation}
which indicates diffusion, but with a diffusion constant 
\begin{equation}
D = \frac{E_0^2\gamma}{2[\hbar^2\gamma ^2 +(\hbar\omega-\alpha)^2]}
=\frac{E_0^2}{2\hbar^2\left[\gamma_0(\frac{\Omega}{\omega_L-\omega_0})^2+
\frac{[(\omega-\omega_B)(\omega_L-\omega_0)]^2}{\gamma_0\Omega^2}  \right]},
\label{76}
\end{equation}
tunable with the external derive of frequency $\omega$. This is one
of the main results of this work. The energy-level spacing $\alpha = F d =\omega_B\hbar$ between the WSL states can be controlled by the imparted acceleration for optical lattice case and by electrostatic field ${\bf E}$ as $ \alpha = e{\bf E}. {\bf a}$  for semiconductor superlattice case.  Thus $\alpha$ and $\omega$ acts as control parameters in an experiment. Diffusion coefficient becomes maximum in the on-resonance case:
\[\omega_c=\frac{\alpha}{\hbar}=\frac{F d}{\hbar}=\omega_B (Bloch~frequency)\] 
with
\begin{equation}
D=\frac{\beta^2}{2\gamma} = \frac{E_0^2}{2\gamma_0\hbar^2}\left[\frac{\omega_L-\omega_0}{\Omega}\right]^2.
\end{equation}
Which is exactly the equation (17), identifying $E_0$ with $V$. 
\section{Discussion}
We have studied some transport properties of cold atoms in an accelerated and harmonically driven optical lattices in the presence of decohering effects due to spontaneous emission. The novelty of the work is a tunable diffusion coefficient and the possibility to control the diffusve transport by external control parameters. We consider a practically important case, in which our system (an atom on an tilted optical lattice) is present in an external AC drive with wavelength longer than lattice period, so as to subtend a strong dipolar matrix element between the neighboring WSL states. We have obtained several interesting results about mean-squared displacement (Figures 4 to 7). Our main result is a tunable diffusion coefficient (equation (28)), which becomes maximum when $\omega_{AC} = \omega_{Bloch} $, (equation (29)), (i.e., the drive frequency is equal to Bloch frequency). Similar effects has been observed for ultra cold Bosonic atoms, i.e., a system of ultracold bosonic atoms in a tilted optical lattice can become superfluid (Mott-insulator to superfluid transition) in response to resonant AC forcing\cite{eck07,eck05,cre06,lig07}. Underlying mechanism is the dynamical disappearance of the energy gap due to static tilt and external AC drive. There is a possibility to study the effect of decoherence on this kind of AC indued flow.

This diffusion maximization of a single atom should be contrasted with 'localization' (dynamical localization of dynamical disorder\cite{dunlup,madison} and Anderson localization of static disorder\cite{ander}). Dynamical localization is related to the suppression of electron transport in a Bloch band driven by an AC electric field, which is explained on the basis of Floquet theory of Bloch band collapse\cite{holthaus}. \emph{Here we report an opposite effect, a kind of 'maximum non-localization'}. The physical mechanism under action appears to be the simultaneous effects of decoherence (due to spontaneous emission) and dynamical disappearance of energy gap between nearby states of WSL, when drive photon energy is equal to the WSL gap energy. Because these are the two main physical effects under action.

For numerical estimation, consider that we have Na atoms (consider very dilute atomic sample so that one can neglect atom-atom interactions) on an optical lattice created by using two counter propagating laser beams of wave length $\lambda = 852 nm$\cite{lig07}. The resulting lattice spacing is $d_L=\lambda/2 = 0.426\mu m$. It is accelerated at a rate $a = d_L\frac{d}{dt}\Delta\nu = 1000 m/sec^2 $. Here $\Delta\nu$ is the difference between the frequencies of two lattice beams. The diffusion will be maximum for $\omega_{AC} = \omega_{Bloch} =\frac{F d}{\hbar}=\frac{m a d_L}{\hbar} \simeq 154 kHz$, and the separation ($\alpha=Fd$) between nearest WSL states will be of the order of $100 peV$. The depth $V_0$ of the resulting periodic potential can be $6 E_{rec} ,\,\,E_{rec} = \frac{\hbar^2\pi^2}{2md_L^2}$. For the estimation of diffusion coefficient at resonance between \emph{WSL and external AC field}, but in the far detuned regime between \emph{atom and optical-lattice laser field}, where only virtual transitions can occur, and one can ignore the internal structure of the atom(adiabatic elimination procedure), we take Rabi frequency $\Omega = 2\mu E_L/\hbar = 2\pi\times34.5\times 10^7 Hz$, detuning $\delta_L= 2\pi\times 5.4\times 10^9 Hz$, rate of spontaneous emission (for sodium atoms) $\gamma_0 = 6.2\times10^7 Hz$\cite{goet96}. With this $D = \frac{E_0^2 d^2}{2\gamma_0\hbar^2}[\frac{\delta_L}{\Omega}]^2 = 1.976\times10^{-6}~Sec$ (in terms of scaled units $d = 1(unit~lattice~spacing), E_0(amplitude~of~transfer~matrix~element) = 1,\hbar = 1$).
\section{Acknowledgments}
I am indebted to Professor N Kumar for critical discussion and constant encouragement.
\section*{Appendix A: No acceleration and no AC Drive}
We take the time(scaled) laplace transform $\tilde{\rho}(p,u,s)=\int_0^\infty e^{-s\tau}\rho(p,u,\tau)d\tau$ of equation (14). and get
\begin{equation}
s\tilde{\rho}(p,u,s)-\rho(p,u,t=0)=\phi(p,u)\tilde{\rho}(p,u,s)+\Gamma\tilde{\chi}(u,s).
\label{12}
\end{equation}
We want to calculate the value of $\rho(p,u,t=0)$. We know that 
\[\rho_{m,n}(t=0)=\langle C^{\ast}_m(t=0)C_n(t=0)\rangle =\delta_{m,0}\delta_{n,0}\]
\begin{eqnarray}
&&\tilde{\rho}(k_1,k_2,t=0) = \sum_{m,n}\rho_{m,n}(t=0) e^{-i m k_1}e^{i n k_2}
=\sum_{m,n}\delta_{m,0}\delta_{n,0} e^{-i m k_1} e^{i n k_2}
\nonumber\\&&\rho(p,u,t=0) = \sum_{m,n}\delta_{m,0}\delta_{n,0}e^{-i m(p+u/2)}
e^{i n(p-u/2)}=1,\nonumber\\
\label{13}
\end{eqnarray}
with this we get
\begin{equation}
\tilde{\rho}(p,u,s) = \frac{1+\Gamma\tilde{\chi}(u,s)}{s-\phi(p,u)}.
\label{14}
\end{equation}
Summing the above equation i.e., equation (32) over p.
\begin{eqnarray}
&&\sum_p\tilde{\rho}(p,u,s) = \sum_p\frac{1+\Gamma\tilde{\chi}(u,s)}{s-\phi(p,u)},\nonumber\\
&&\frac{1}{2\pi}\int_{-\pi}^{\pi}\tilde{\rho}(p,u,s)dp = \frac{[1+\Gamma\tilde{\chi}(u,s)]}{2\pi}\int_{-\pi}^{\pi}\frac{dp}{s-\phi(p,u)}
\label{15}.
\end{eqnarray}
By re-arrangements we get
\begin{equation}
\tilde{\chi}(u,s) = \frac{I}{1-I\Gamma}\;\;,\;\; I=\frac{1}{2\pi}\int_{-\pi}^{\pi}
\frac{dp}{s-\phi(p,u)}.
\label{16}
\end{equation}
Now, we want to find the mean and mean-squared displacement, as we know
\begin{eqnarray}
&&\tilde{\chi}(u,s)=\frac{1}{2\pi}\int_{-\pi}^{\pi}\tilde{\rho}(q,u,s)dq=\nonumber\\
&&\frac{1}{2\pi}\int_{-\pi}^{\pi}\sum_{m,n}\tilde{\rho}_{m,n}(s)e^{-im(p+u/2)}
e^{in(p-u/2)}dp\nonumber\\
&&=\frac{1}{2\pi}\int_{-\pi}^{\pi}\sum_{m,n}\tilde{\rho}_{m,n}(s)e^{-i(m+n)
\frac{u}{2}}e^{i(n-m)p}dp\nonumber\\
&&=\sum_{m,n}\delta_{m,n}\tilde{\rho}_{m,n}(s)e^{-i(m+n)\frac{u}{2}}
=\sum_{n}\tilde{\rho}_{n,n}(s)e^{-i n u}\nonumber\\
&&\delta=\frac{1}{2\pi}\int_{-\pi}^\pi e^{i(n-m)p}dp.
\label{17}
\end{eqnarray}
The mean displacement in $s$-domain is given by $\sum_n n \tilde{\rho}_{n,n}(s)$, noting that\[\frac{\partial\tilde{\chi}(u,s)}{\partial u}=-i\sum_n n\tilde{\rho}_{n,n}(s)e^{-inu}.\]
We have
\begin{equation}
\langle\tilde{x}(s)\rangle =i\left[\frac{\partial\tilde{\chi}(u,s)}{\partial u}\right]_{u=0}.
\label{18}
\end{equation}
Similarly, mean-squared displacement is given by
\begin{equation}
\langle\tilde{x^2}(s)\rangle =\sum_n n^2 \tilde{\rho}_{n,n}(s)=-\left[\frac{\partial^2
\tilde{\chi}(u,s)}{\partial u^2}\right]_{u=0}.
\label{19}
\end{equation}
Now differentiating equation (34), w. r. t. $u$ and finding the differentials of the integral $I$ at $u=0$, using equation (36) and equation (37), we obtain
\begin{eqnarray}
\langle\tilde{x}(s)\rangle = 0 \;,\nonumber\\
\langle{x}(t)\rangle=0.
\label{20}
\end{eqnarray}
\begin{equation}
\langle\tilde{x^2}(s)\rangle = \frac{1}{s^2(s+\Gamma)}.
\end{equation}
After inversion, we get the mean-squared displacement (equation (15))
\section*{Appendix B: Accelerated lattice in an external AC drive}
\begin{eqnarray}
&&\frac{\partial \varrho(p,u,\tau)}{\partial \tau} = [2i \sin(p+\Delta\tau)\sin(u/2)-\Gamma]
\times\varrho(p,u,\tau)\nonumber\\
&&+\frac{\Gamma}{2\pi}\int_{-\pi}^{\pi}\varrho(p-q,u,\tau)dq.
\label{60}
\end{eqnarray}
The solution of first order P.D.E (equation (40)) is
\begin{eqnarray}
&&\varrho(p,u,\tau)=\Gamma e^{-\varphi(p,u,\tau)}\int e^{\varphi(p,u,\tau)}\bar
{\chi}(u,\tau)d\tau \nonumber\\
&& + C_1(p,u)e^{-\varphi (p,u,\tau)}.
\label{63}
\end{eqnarray}
In the above, we have defined
\begin{equation}
\bar{\chi}(u,\tau)=\frac{1}{2\pi}\int_{-\pi}^{\pi}\varrho(p-q,u,\tau)dq = 
\frac{1}{2\pi}\int_{-\pi}^{\pi}\varrho(q,u,\tau)dq, 
\label{61}
\end{equation}
\begin{eqnarray}
&&\varphi (p,u,\tau) = -\int [2 i \sin(p+\Delta\tau)\sin(u/2)- \Gamma]d\tau \nonumber\\
&&= \frac{2 i}{\Delta}\cos(p + \Delta\tau)\sin(u/2) + \Gamma\tau.
\label{62}
\end{eqnarray}
Summing over p,
\begin{eqnarray}
&&\sum_p \varrho(p,u,\tau)=\Gamma \sum_p e^{-\varphi (p,u,\tau)}\int e^{\varphi (p,u,\tau)}\bar{\chi}(u,\tau)d\tau \nonumber\\
&&+ \sum_p C_1(p,u)e^{-\varphi (p,u,\tau)}.
\label{64}
\end{eqnarray}
\begin{eqnarray}
&&\frac{1}{2\pi}\int_{-\pi}^{\pi}dp \varrho(p,u,\tau)=\frac{\Gamma}{2\pi}\int_{-\pi}^{\pi}dp 
e^{-\varphi (p,u,\tau)}\nonumber\\
&&\times\int e^{\varphi (p,u,\tau)}
\bar{\chi}(u,\tau)d\tau 
 + \frac{1}{2\pi}\int_{-\pi}^{\pi}dp C_1(p,u)e^{-\varphi (p,u,\tau)}.\nonumber\\
\label{65}
\end{eqnarray}
\begin{eqnarray}
&&\bar{\chi}(u,\tau)=\frac{\Gamma}{2\pi}\int_{-\pi}^{\pi} e^{-\varphi (p,u,\tau)}I_4 dp \nonumber\\
&&+ \frac{1}{2\pi}\int_{-\pi}^{\pi} C_1(p,u)e^{-\varphi (p,u,\tau)}dp,I_4 = 
\int e^{\varphi (p,u,\tau)}\bar{\chi}(u,\tau)d\tau.\nonumber\\
\label{66}
\end{eqnarray} 
To calculate $C_1(p,u)$, we put $\tau = 0$ in equation (46), and use the initial condition equation (7), i.e.,$\bar{\rho}_{mn}(t = 0) =\rho_{mn}(t = 0) = \delta_{m0}\delta_{n0}$,
we have
\begin{eqnarray}
&&\varrho(p,u,\tau=0)=\sum_{m,n}\bar{\rho}_{mn}(0)e^{-i m(p-u/2)}e^{i n (p+u/2)}
\nonumber\\
&&=\sum_{m,n}\delta_{m0}\delta_{n0}e^{-i m(p-u/2)}e^{i n(p+u/2)} = 1.
\label{67}
\end{eqnarray}
So,
\begin{equation}
C_1(p,u)=e^{\varphi(p,u,0)}-\Gamma I_{4\tau},I_{4\tau}=\left[\int e^{\varphi (p,u,\tau)}
\bar{\chi}(u,\tau)d\tau\right]_{\tau=0}.
\label{68}
\end{equation}
Equations (46), and (48) gives
\begin{eqnarray}
&&\bar{\chi}(u,\tau)=\frac{\Gamma}{2\pi}\int_{-\pi}^{\pi} e^{-\varphi (p,u,\tau)}I_4 dp + \frac{1}{2\pi}\int_{-\pi}^{\pi} [e^{\varphi(p,u,0)}
-\Gamma I_{4\tau}]\nonumber\\
&&\times e^{-\varphi (p,u,\tau)}dp ,I_4 = \int e^{\varphi (p,u,\tau)}\bar{\chi}(u,\tau)d\tau,\nonumber\\
&&I_{4\tau}=\left[\int e^{\varphi (p,u,\tau)}\bar{\chi}(u,\tau)d\tau\right]_{\tau=0},\nonumber\\
&&\varphi (p,u,\tau) = \frac{2 i}{\Delta}\cos(p + \Delta\tau)\sin(u/2) + \Gamma\tau.
\label{69}
\end{eqnarray} 
As we have, $\left< x(\tau)\right> = i\frac{\partial \bar{\chi}(u,\tau)}{\partial u}$. In order to calculate the mean displacement, we differentiate the above integral equation for reduced-transformed density matrix equation (49)  w. r. t. $u$ and set $u=0$. Noting that
\begin{eqnarray}
&&\bar{\chi}(u,\tau)=\frac{1}{2\pi}\int_{-\pi}^{\pi}\varrho(q,u,\tau)dq\nonumber\\ &&=\frac{1}{2\pi}\int_{-\pi}^{\pi}\sum_{m,n}\rho_{mn}(\tau)e^{i\delta(m-n)\tau}
e^{-i m(p-u/2)}e^{i n (p+u/2)}dp,
\nonumber\\
&&\bar{\chi}(0,\tau)=\frac{1}{2\pi}\int_{-\pi}^{\pi} \sum_{m,n}\rho_{mn}(\tau)e^{i\delta(m-n)\tau}e^{i(n-m)p}dp\nonumber\\
&& = \sum_{m,n}\delta_{mn}\rho_{mn}(\tau)e^{i\delta(m-n)\tau}=1.
\label{70}
\end{eqnarray}
We finally obtain
\begin{equation}
\langle x(\tau)\rangle = \left[\frac{\partial \bar{\chi}(u,\tau)}{\partial u}\right]_{u=0} = 0.
\label{71}
\end{equation}
To calculate mean-squared displacement 
$=-\left[\frac{{\partial}^2\bar{\chi}(u,\tau)}{\partial u^2}\right]_{u=0}$, we solve equation (49) by doubly differentiating it w. r .t. $u$ and then 
setting $u=0$ to get
\begin{eqnarray}
&&\left[\frac{{\partial}^2\bar{\chi}(u,\tau)}{\partial u^2}\right]_{u=0} = 
\Gamma e^{-\Gamma\tau}\int e^{\Gamma\tau}\left[\frac{{\partial}^2\bar{\chi}(u,\tau)}
{\partial u^2}\right]_{u=0}d\tau\nonumber\\ &&-\frac{1}{(\Delta^2+\Gamma^2)}(1-e^{-\Gamma\tau}\cos \Delta\tau)
+\frac{\Gamma}{\Delta(\Delta^2+\Gamma^2)}e^{-\Gamma\tau}\sin \Delta\tau\nonumber\\
&&-2\pi\left[\int e^{\Gamma\tau}\left[\frac{{\partial}^2\bar{\chi}(u,\tau)}{\partial u^2}\right]_{u=0} d\tau\right]_{\tau = 0} e^{-\Gamma\tau}.
\label{72}
\end{eqnarray}
Equation (52) is solved by Laplace Transform method.
In equation (52), we define $e^{\Gamma\tau}\left[\frac{{\partial}^2\bar{\chi}(u,\tau)}{\partial u^2}\right]_{u=0} = f(\tau)$. With this, $f(\tau)$ takes the following form
\begin{equation}
f(\tau)=\Gamma \int f(\tau)d\tau - \frac{e^{\Gamma\tau}}{\Delta^2 + \Gamma^2} +
\frac{\cos\Delta\tau}{\Delta^2 + \Gamma^2}
+ \frac{\sin \Delta\tau}{\Delta(\Delta^2+\Gamma^2)}+constant.
\end{equation}
Differentiating the above integral equation, we get
\begin{equation}
\frac{df(\tau)}{d\tau} = \Gamma f(\tau) - \frac{\Gamma e^{\Gamma\tau}}{\Delta^2 + \Gamma^2} +
\frac{\Gamma \cos\Delta\tau}{\Delta^2 + \Gamma^2} - \frac{\Delta\sin\Delta\tau}
{(\Delta^2+\Gamma^2)}.
\end{equation}
With the initial condition $f(\tau =0) = 0$, i.e., mean-squared displacement is zero at time $\tau = 0$, the above equation can be readily solved to get the mean-squared displacement as given in equation (25).

\end{document}